\documentclass[twocolumn,showpacs,preprintnumbers]{revtex4}%
\usepackage{amsfonts}
\usepackage{amsmath}
\usepackage{graphicx}
\usepackage{dcolumn}
\usepackage{bm}
\usepackage{amssymb}%

\usepackage{soul}

\begin{document}
\title{Chiral density wave in nuclear matter}
\author{Achim Heinz, Francesco Giacosa, and Dirk H.\ Rischke}
\affiliation{Institute for Theoretical Physics, Goethe University,
Max-von-Laue-Str.\ 1, D--60438 Frankfurt am Main, Germany }

\begin{abstract}
Inspired by recent work on inhomogeneous chiral condensation in
cold, dense quark matter within models featuring quark degrees
of freedom, we investigate the chiral density-wave solution in
nuclear matter at zero temperature and nonvanishing baryon number
density in the framework of the so-called
extended linear sigma model (eLSM). The eLSM is an effective model
for the strong interaction based on the global chiral 
symmetry of quantum chromodynamics (QCD). It contains scalar, pseudoscalar,
vector, and axial-vector mesons as well as
baryons. In the latter sector, the nucleon and its chiral partner are
introduced as parity doublets in the mirror assignment. 
The eLSM simultaneously provides a good description of hadrons in vacuum 
as well as nuclear matter ground-state properties. We find
that an inhomogeneous phase in the form of a chiral density wave is
realized, but only for
densities larger than $2.4\rho_{0}$, where $\rho_{0}$ is the nuclear matter
ground-state density.

\end{abstract}

\maketitle



\emph{Introduction:} The spontaneous breaking of chiral symmetry in the QCD
vacuum is a nonperturbative phenomenon which has to be reflected
in low-energy hadronic theories, see e.g.\ Refs.\
\cite{meissner,geffen,lee}. The order parameter of chiral symmetry
breaking is the chiral condensate, denoted as $\langle\bar{q} q
\rangle \sim \langle\sigma\rangle$,
which contributes to hadronic masses and is responsible for
the mass splitting of so-called chiral partners, i.e., hadrons with
the same quantum numbers except for parity and G-parity.

At sufficiently large temperature and density, it is expected that the
spontaneously broken chiral symmetry is (at least partially) restored.
Lattice-QCD calculations \cite{Aoki:2006br,Cheng:2006qk} show
that, for values of the quark masses realized in nature, 
this so-called chiral transition is cross-over along the temperature axis of
the QCD phase diagram. Along the
density axis, lattice-QCD calculations are not yet available (for
realistic quark masses), but phenomenological models 
\cite{compactstars,Mishustin:1993ub} indicate that chiral symmetry
restoration may occur through a first-order phase transition.

An interesting possibility is that the effective potential is minimized
by an order parameter which varies as a function of spatial coordinate.
Such inhomogeneous phases were already suggested in the pioneering works of Ref.\
\cite{Baym:1973zk,Sawyer:1973fv,Campbell:1974qt,Campbell:1974qu,Dashen:1974ff,Baym:1975tm,
Migdal:1978az,Baym:1980jx,Dickhoff:1981wk,Kolehmainen:1982jn,Baym:1982ca,Akmal:1997ft}.
In particular, the chiral condensate may assume the form of 
the so-called chiral-density wave where not only the chiral condensate
$\langle\sigma\rangle$ but also the expectation value of the neutral pion field
is non-vanishing, $\left\langle \pi^{3}\right\rangle \neq0.$ 
However, the problem of the aforementioned approaches
was that, without nucleon-nucleon tensor forces, inhomogeneous chiral
condensation took place already in the nuclear matter ground state, in 
contradiction to experimental findings.

More recently, inhomogeneous phases were studied in the
framework of the (1+1)--dimensional Gross-Niveau model
\cite{Schnetz:2004vr,Wagner:2007he}, where it was indeed found 
that a spatially varying order parameter minimizes the effective potential at high
density. In Ref.\ \cite{Kojo:2009ha,quarkyonic}, the authors coined the
phrase ``quarkyonic matter'' for an inhomogeneous phase at high
density, where the
chiral density-wave solution is realized within QCD in the large--$N_c$ limit.
Inhomogeneous phases were also
investigated in Refs.\ \cite{Nickel:2009wj,Carignano:2012sx,
Broniowski:2011ef,ebert,Ma:2013vga,Tatsumi:2004dx} in the
framework of the Nambu--Jona-Lasinio as well as the quark-meson model.

In this work, we re-investigate the question whether
inhomogeneous condensation at nonzero density occurs in a model solely
based on the degrees of freedom of the QCD vacuum, i.e., hadrons.
We employ dilatation invariance and chiral symmetry of QCD, and include
scalar, pseudoscalar, vector, and axial-vector mesons, as well as
baryons and their chiral partners. This approach,
developed in Refs.\ \cite{denis,dick,gallas1} and denoted as extended
Linear Sigma Model (eLSM), successfully describes hadron vacuum phenomenology
both in the meson \cite{denis,dick} and baryon \cite{gallas1} sectors. In the
latter, the nucleon and its chiral partner are treated in the
mirror assignment \cite{DeTar:1988kn,jido}, in which a chirally
invariant mass term exists [see also Refs.\
\cite{zschiesche,sasakimishustin,peng,gallas2,gallasnew} and refs.\ therein].

The chiral condensate $\langle\sigma\rangle$ is the expectation value of
the $\sigma$ field which is the chiral partner of the pion $\vec{\pi}.$
In the framework of the eLSM the resonance corresponding to the
$\sigma$ field
is not the lightest scalar resonance $f_{0}(500)$ (as proposed in older versions
of the $\sigma$ model), but the heavier state $f_{0}(1370).$ This
result is in agreement with a variety of studies of low-energy QCD, e.g.\
Refs.\ \cite{amslerrev,refs} and refs.\ therein. In Ref.\ \cite{pelaez}, 
the $f_0(500)$ state is interpreted as a resonance in the pion-pion scattering
continuum. Other works \cite{jaffe,fariborz,tqmix} favor an
interpretation of $f_{0}(500)$ as a tetraquark state. This fact has an
important consequence for studies at nonzero density; namely, when using a
chiral linear sigma model, both resonances $f_{0}(500)$ and
$f_{0}(1370)$ should be taken into account, the former being the
lightest scalar state and the latter being
an excitation of the chiral condensate. This was, for instance,
done in Ref.\ \cite{gallas2} where, in the framework of the eLSM, 
the resonance $f_{0}(500)$ was coupled in a
chirally invariant manner to nucleons and their chiral partners. 
An important result of this study was that the
nuclear matter ground-state properties (i.e., density, binding energy,
and compressibility) could be successfully described. 
In the mean-field approximation, and assuming homogeneous condensates,
Ref.\ \cite{gallas2} reports the onset of a first-order phase
transition at a density of about $2.5\rho_{0}$. 
The important role of both aforementioned scalar resonances has
been also investigated at nonzero temperature in the framework of a simplified
version of the eLSM in Ref.\ \cite{achim}. Interestingly, the necessity to
include both scalar-isoscalar states $f_{0}(500)$ and
$f_{0}(1370)$ has been also shown in the framework of the Bonn nucleon-nucleon
potential \cite{ml}.

The main question of the present work is whether inhomogeneous 
condensation takes place within the eLSM. For simplicity, we restrict
ourselves to a spatial dependence of the condensate, which
is of the type of the chiral density wave: $\langle\sigma
\rangle\sim\cos(2fx)$, $\langle\pi^{3}\rangle\sim\sin(2fx)$. 
We will show, by using the parameters determined in Ref.\
\cite{gallas2} to describe nuclear matter ground-state properties 
(and thus having no additional free parameters), that 
nuclear matter in the ground state is still a (homogeneous) liquid.
Chiral condensation remains homogeneous up to a chemical
potential of $973$ MeV (corresponding to a density of $2.4\rho_{0}$),
followed by an inhomogeneous phase with a chiral density wavelength 
$\pi/f$ of approximately $1.5$ fm.

\emph{The model: }In the two-flavor case, $N_{f}=2,$ the scalar and
pseudoscalar sectors are described by the matrix
\[
\Phi=(\sigma+\imath\eta_{N})t_{0}+(\vec{a}_{0}+\imath\vec{\pi})\cdot\vec{t}~,
\]
where $\vec{t}=\vec{\tau}/2$, with the vector of Pauli matrices $\vec{\tau}$, and
$t_{0}=\mathbf{1}_{2}/2.$ The vector and axial-vector mesons enter via the
matrices
\[
V^{\mu}=\omega^{\mu}t_{0}+\vec{\rho}^{\mu}\cdot\vec{t}~,\;\;A^{\mu}=f_{1}^{\mu
}t_{0}+\vec{a}_{1}^{\mu}\cdot\vec{t}~,
\]
from which the left-handed and right-handed vector fields are defined as
$R^{\mu}\equiv V^{\mu}-A^{\mu}$, $L^{\mu}\equiv
V^{\mu}+A^{\mu}$. Under
the chiral group $SU(2)_{R}\times SU(2)_{L}$ the fields transform as
$\Phi\rightarrow U_{L}\Phi U_{R}^{\dagger}$, $R^{\mu}\rightarrow U_{R}R^{\mu
}U_{R}^{\dagger}$, and $L^{\mu}\rightarrow U_{L}L^{\mu}U_{L}^{\dagger}$. The
identification of mesons with particles listed in Ref.\ \cite{PDG} is as
follows: the fields $\vec{\pi}$ and $\eta_{N}$ correspond to the pion and the
nonstrange part of the $\eta$ meson, $\eta_{N}\equiv(\overline
{u}u+\overline{d}d)/\sqrt{2}$. The fields $\omega^{\mu}$ and $\vec{\rho}^{\mu
}$ represent the vector mesons $\omega(782)$ and $\rho(770)$, and the fields
$f_{1}^{\mu}$ and $\vec{a}_{1}^{\mu}$ the axial-vector mesons
$f_{1}(1285)$ and $a_{1}(1260)$. The scalar fields $\sigma$ and
$\vec{a}_{0}$ fields are identified with $f_{0}(1370)$ and
$a_{0}(1450)$, respectively. The
chiral condensate $\phi=\left\langle \sigma\right\rangle =Zf_{\pi}$ emerges
upon spontaneous chiral symmetry breaking in the mesonic sector, where
$f_{\pi}\simeq 92.4$ MeV is the pion decay constant and $Z \simeq 1.67$ 
is the wave-function renormalization constant of the pseudoscalar
fields \cite{denis}.

In the present work, besides ordinary quarkonium mesons, also the lightest
resonance $f_{0}(500)$ is introduced \cite{gallas2}.  For $N_f=2$,
it does not matter whether we interpret $f_{0}(500)$ as
(predominantly) a tetraquark field or as a pion-pion resonance; differences
in the coupling to other fields occur, however, for $N_f \geq 3$. 
The bare $f_0(500)$ field is denoted as
$\chi$. It is a singlet under chiral transformation and is
coupled to mesons following Refs.\ \cite{tqmix,achim}.

We first consider the mesonic part of the eLSM Lagrangian and keep
only terms involving fields which will eventually condense, i.e.,
$\sigma$, $\pi = \pi^{3}$ , $\omega_{\mu}$, and $\chi$ 
[for the full Lagrangian, see Ref.\ \cite{denis}]:
\begin{align}
\mathcal{L}_{\text{mes}}  &  =\frac{1}{2}\partial_{\mu}\sigma\partial^{\mu
}\sigma +\frac{1}{2}\partial_{\mu}\pi\partial^{\mu}\pi\nonumber\\
&  -\frac{1}{4}(\partial_{\mu}\omega_{\nu}-\partial_{\nu}\omega_{\mu}%
)^{2}+\frac{1}{2}\partial_{\mu}\chi\partial^{\mu}\chi\nonumber\\
& +\frac{1}{2}m^{2}(\sigma^{2}+\pi^{2})
-\frac{\lambda}{4}(\sigma^{2}+\pi^{2})^{2}
+\varepsilon\sigma\nonumber\\
&  +\frac{1}{2}m_{\omega}^{2}\omega_{0}^{2}
-\frac{1}{2}m_{\chi}^{2}\chi^{2}+g\chi(\sigma^{2}+\pi^{2})\text{ .}
\label{vmes}%
\end{align}
The following numerical values are used \cite{denis,tqmix}:
$m^{2}=(896.3$ MeV)$^{2}$, $\lambda=35.05,$  $\varepsilon
=1.054\cdot10^{6}$ MeV$^{3},$ $g=438$ MeV, $m_{\omega}=782$ MeV, and $m_{\chi}=611$
MeV. As a consequence of this choice of parameters, 
$m_{\sigma}=1295$ MeV and $\sigma$ is identified with $f_{0}(1370).$

We now make the following Ansatz for the condensates, 
which is of the form of a chiral density wave:
\begin{align}
\langle\sigma\rangle &  =\phi\cos(2fx)\text{ , }\langle\pi\rangle
=\phi\sin(2fx)\text{ ,}\label{cdw1}
\end{align}
In the limit $f \rightarrow 0$ we obtain the usual homogeneous
condensation which is
realized in the vacuum and, as we shall see, for low densities.
Besides these
inhomogeneous condensates, also $\chi$ and $\omega_{0}$
develop nonvanishing condensates which are, however, homogeneous: $\langle
\chi\rangle=\bar{\chi}$ and $\langle\omega_{0}\rangle=\bar{\omega}_{0}$.

In mean-field approximation, we insert 
the Ansatz (\ref{cdw1}) into
the mesonic part of the Lagrangian (\ref{vmes}) and obtain the
tree-level potential
\begin{align}
U_{\text{mes}}^{\text{mean-field}}  &  =  2f^{2}\phi
^{2}+\frac{\lambda}{4}\phi^{4}-\frac{1}{2}m^{2}\phi^{2}-\varepsilon\phi
\cos(2fx) \nonumber\\
&  -\frac{1}{2}m_{\omega}^{2}\bar{\omega}_{0}^{2}+\frac{1}{2}m_{\chi
}^{2}\bar{\chi}^{2}-g\bar{\chi}\phi^{2}  \text{ .}%
\end{align}
Note that, in a spatial volume $V > (\pi/f)^3$, 
the spatially dependent term $\sim \cos (2 f x)$ averages
to zero for any nonzero value of $f$.

We now turn to the baryonic sector where we introduce two doublets $\Psi_{1}$ and
$\Psi_{2}$ transforming according to the mirror assignment:
\begin{align}
\Psi_{1,R}  &  \rightarrow U_{R}~\Psi_{1,R}~,~~~~~~\Psi_{1,L}\rightarrow
U_{L}~\Psi_{1,L}~,\\
\Psi_{2,R}  &  \rightarrow U_{L}~\Psi_{2,R}~,~~~~~~\Psi_{2,L}\rightarrow
U_{R}~\Psi_{2,L}~.
\end{align}
The mirror assignment has the consequence that the
following mass term is allowed by chiral symmetry 
\cite{DeTar:1988kn,jido,gallas1,gallasnew}:
\begin{equation}
m_{0}\left(  \bar{\Psi}_{1,L}\Psi_{2,R}-\bar{\Psi}_{1,R}
\Psi_{2,L}-\bar{\Psi}_{2,L}\Psi_{1,R}+\bar{\Psi}_{2,R}
\Psi_{1,L}\right)~.  ~ \label{m0}%
\end{equation}
However, in order to preserve dilatation invariance for the baryon sector, the
constant $m_{0}$ should emerge upon condensation of scalar fields.
These fields can be assigned as a scalar-isoscalar glueball or a
tetraquark (molecular) state \cite{gallas1,gallas2}; 
here we only consider for simplicity the latter possibility, i.e., the
coupling of baryons to the field $\chi$.
The resulting baryonic Lagrangian in which only the mesons
$\sigma,$ $\omega,$ and $\chi$ are retained reads:
\begin{align}
  \mathcal{L}_{\text{bar}} & = \overline{\Psi}_{1}\imath\gamma_{\mu}\partial^{\mu
}\Psi_{1}+\overline{\Psi}_{2}\imath\gamma_{\mu}\partial^{\mu}\Psi_{2}
\nonumber \\
& -\frac{\widehat{g}_{1}}{2}\overline{\Psi}_{1}(\sigma+ \imath \gamma_5
\tau^3 \pi)\Psi_{1}
  -\frac{\widehat{g}_{2}}{2}\overline{\Psi}_{2}(\sigma- \imath \gamma_5
\tau^3 \pi)\Psi_{2}\nonumber \\
& -g_{\omega}\overline{\Psi}_{1}\imath\gamma_{\mu}\omega^{\mu}\Psi_{1}-g_{\omega}
\overline{\Psi}_{2}\imath\gamma_{\mu}\omega^{\mu}\Psi_{2}\nonumber\\
&  -a\chi(\overline{\Psi}_{2}\gamma_{5}\Psi_{1}-\overline{\Psi}_{1}\gamma
_{5}\Psi_{2})\text{ .} \label{lageff}%
\end{align}
where the parameters $\hat{g}_{1},$ $\hat{g}_{2},$ $g_\omega$, and $a$
are dimensionless, in accordance with dilatation invariance.
The mass term $m_{0}$ of Eq.\ (\ref{m0}) emerges upon condensation of the
tetraquark field: $m_{0}=a\bar{\chi}.$ The choice of parameters which
reproduces nuclear matter ground-state properties is
\cite{gallas2}: $\hat{g}_{1}=10.80,$ $\hat{g}_{2}=18.53,$ $g_{\omega}
=5.04,$ and $a=17.92$ (which implies $m_{0}=500$ MeV). 
For these values the compressibility is $K = 197.0$ MeV. 
This is consistent with the range $200 -300$ MeV
quoted in Refs.\ \cite{Youngblood:2004fe,Hartnack:2007zz}. In particular, our
result is in reasonable agreement with the value $K = 235 \pm 14$ MeV obtained in Ref.\
\cite{Youngblood:1999zza} by studying inelastic scattering of $\alpha$
particles off nuclei. 

When studying the chiral density-wave Ansatz of Eq.\ (\ref{cdw1}) it is useful to
make the following field redefinitions of the baryon fields \cite{ebert}:
\[
\Psi_{1}\rightarrow\Psi_{1}\exp\left(  -\imath\gamma_{5}\tau^{3}fx\right)
\text{~, }\Psi_{2}\rightarrow\Psi_{2}\exp\left(  +\imath\gamma_{5}\tau
^{3}fx\right)  ~,
\]
thanks to which the explicit spatial coordinate dependence transforms into
a momentum dependence. In the effective potential,
we treat the meson fields in the mean-field
approximation, while the fermions are integrated out. In the no-sea
approximation for the fermions, the
effective potential then reads
\begin{align}
U&_{\rm eff}(\phi,\bar{\chi},\bar{\omega}_0,f) = \nonumber \\
& \sum_{k=1}^{4}\int\frac{2d^{3}p}{(2\pi)^{3}}[E_{k}%
(p)-\mu^{\ast}]\Theta[\mu^{\ast}-E_{k}(p)] 
+U_{\text{mes}}^{\text{mean-field}}\;,
\end{align}
where $\mu^{\ast}=\mu-g_{\omega}\bar{\omega}_{0}$ and $E_{k}(p)=\sqrt
{p^{2}+\bar{m}_{k}(p_{x})^{2}}$. In the case of homogeneous
condensation there are only two (two-fold degenerate) energy eigenstates,
corresponding to the nucleon and its chiral partner, while for inhomogeneous
condensation the degeneracy is lifted and four different energy
eigenstates emerge. The values for $E_k(p)$ are calculated numerically as solutions of
characteristic polynomials.

Note that, within the mean-field approximation,
the baryon density remains constant over space, even if sigma and
pion fields are inhomogeneous. This is in contrast to the case in which only the neutral
pion condenses, in which also space-like modulations of the nucleon fields are
obtained \cite{Takatsuka:1978ku}. However, our result could change when additional
contributions, such as the exchange of (pseudo)tensor mesons, are taken into
account.

\emph{Results:} The effective potential is numerically minimized with
respect to $\phi$, $\bar{\chi}$, and $f$ and maximized 
with respect to $\bar{\omega}_{0}$, respectively.
In Fig.\ 1 we show the effective potential as a function of 
$\phi$ (at the extrema of $\bar{\chi}$, $\bar{\omega}_0$, and $f$)
for the chemical potential $\mu=923$~MeV, which corresponds to
the nuclear matter ground state. There are two degenerate global minima, one for $\phi
=154.3$~MeV corresponding to the vacuum, and one for $\phi=149.5$~MeV
corresponding to the nuclear matter ground state, respectively. 
The decrease of the chiral condensate as compared to the vacuum is very small, which is a
consequence of the pseudoscalar wave-function renormalization $Z=1.67>1$ 
[and thus, indirectly, of the presence of (axial-)vector
mesons; for results where (axial-)vector mesons were not taken into account,
see Ref.\ \cite{achimproc}]. Moreover, we also notice the presence
of a local minimum at $\phi=38.3$~MeV, which corresponds to inhomogeneous
condensation. For increasing $\mu$ the position of this minimum
changes only slightly with $\phi$, but it eventually
becomes the global minimum and thus the thermodynamically realized state.

\begin{figure}
[ptb]
\begin{center}
\includegraphics[
height=1.8663in,
width=3.339in
]%
{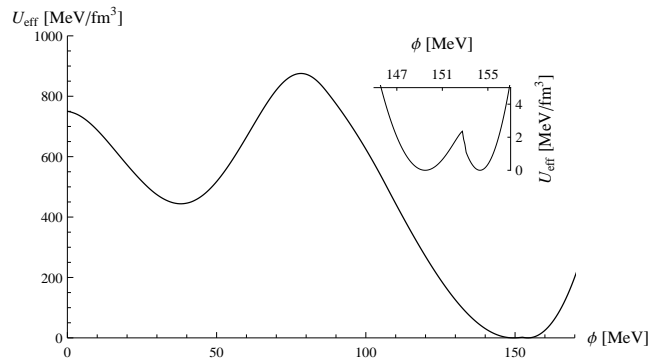}%
\caption{The effective potential $U_{\rm eff}$ 
as a function of $\phi$ for $\mu_{B}=923~\text{MeV}$, at the
extrema for $\bar{\chi},\bar{\omega}_0,$ and $f$.
There are three minima for $\phi=154.4~\text{MeV}$, $\phi=149.5~\text{MeV,
and}$ $\phi=38.3~\text{MeV. }$They correspond to the vacuum, the
nuclear matter
ground state,
and the chiral density-wave state, respectively. }%
\end{center}
\end{figure}
%

\begin{figure}
[ptb]
\begin{center}
\includegraphics[
height=2.0029in,
width=3.3676in
]%
{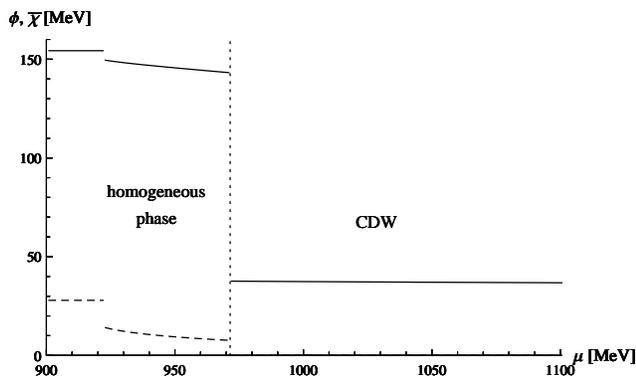}%
\caption{The condensates $\phi$ (solid line) and $\bar{\chi}$ (dashed line) are
plotted as functions of $\mu$. At $\mu=923~\text{MeV}$ a
first-order phase transition to the nuclear matter ground state
takes place. Above $\mu>973~\text{MeV}$ the chiral density wave
becomes the thermodynamically realized
state. In this regime the condensate $\bar{\chi}$ is very small (but
nonzero).}
\end{center}
\end{figure}

%

\begin{figure}
[ptb]
\begin{center}
\includegraphics[
height=1.9545in,
width=3.1237in
]%
{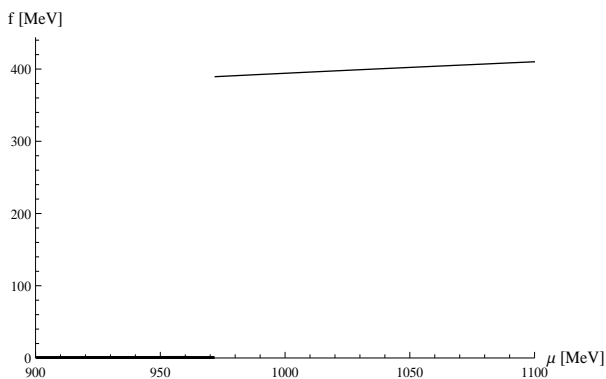}%
\caption{The parameter $f$ as a function of $\mu$.}%
\end{center}
\end{figure}

In Fig.\ 2 the condensates $\phi$ and $\bar{\chi}$ are shown as functions of
$\mu$. For $\mu=923~\text{MeV}$ a first-order phase transition to the nuclear
matter ground state takes place. Both condensates drop and then
further decrease slowly for
increasing $\mu$. At $\mu=973~\text{MeV}$ a transition to the inhomogeneous
phase occurs. The condensate $\bar{\chi}$ drops to (almost) zero 
and the chiral condensate $\phi$
to the value $\phi=37.6~\text{MeV}$. For larger $\mu$ the condensates
$\bar{\chi}$ and $\phi$ change very slowly. Note that $\phi$ does not vanish, thus
chiral symmetry is not completely restored. In terms of density,
the onset of inhomogeneous condensation is at $2.4\rho_{0}$; this density
is still small enough such that a hadronic description of the system
is valid. Then a mixed phase, where homogeneous matter and the chiral
density-wave phase coexist, is realized between $2.4\rho_{0}$ and $10.4\rho_{0}$. 
For densities slightly above $2.4\rho_{0}$ matter is mostly homogeneous and only small
regions (bubbles) with inhomogeneous condensation inside are present. The ratio of
inhomogeneous to homogeneous matter increases with density and at
$10.4\rho_{0}$ only the chiral density-wave phase left. The value $10.4\rho_{0}$ is too large to
trust in a hadronic description, thus we are led to believe that somewhere in the mixed phase a
transition to quark matter should occur, which is, however, not part of our model. 

The onset of the chiral density-wave phase can be best shown plotting the behavior of
the parameter $f$, see Fig.\ 3: $f$ vanishes for small $\mu,$ but jumps to a
nonzero value ($f=389.5$ MeV) at the critical value $\mu=973~$MeV and then
slightly increases for increasing $\mu$. This is the chiral density-wave phase, with a
one-dimensional harmonic modulation with a wavelength of about 1.5 fm. 

Quite remarkably, a similar picture has been obtained in Ref.\
\cite{Ma:2013vga} in the framework of skyrmion matter, in which
inhomogeneous condensation is realized at $2\rho_{0}$ and $m_{0}\sim600$ MeV.
In our case, we have found that decreasing $m_{0}$ leads to an unphysical
inhomogeneous nuclear matter ground state. It would be interesting to see if
this property also occurs in the context of the skyrmion model.

\emph{Conclusions}: We have shown that in the framework of a chiral hadronic
approach (the eLSM) an inhomogeneous chiral condensate becomes favoured at large
density. Using a set of parameters which allows for a correct description
of the nuclear matter ground-state properties, we find that the onset of the
inhomogeneous phase occurs at $2.4\rho_{0}.$ To our knowledge, our result
is the first demonstration of inhomogeneous chiral condensation in an approach
capable of describing vacuum as well as nuclear matter properties. In
future studies one should go beyond the chiral density-wave Ansatz 
and perform a more detailed
analysis of the phases realized in the system. At large densities,
one cannot avoid to include strange degrees of freedom.

\bigskip

\textbf{Acknowledgments:} The authors acknowledge very useful discussions with
I.\ Mishustin, A.\ Schmitt, M.\ Wagner, and J.\ Wambach. A.H.\ acknowledges support from H-QM and
HGS-HIRe. F.G.\ thanks the Foundation Polytechnical Society Frankfurt am Main
for support through an Educator fellowship.

\end{document}